\documentclass[apj]{emulateapj}
\usepackage{graphicx}
\setkeys{Gin}{width=0.9\textwidth}

\newcommand{\kpc}{{\rm kpc}}
\newcommand{\lsun}{{\rm\,L_\odot}}
\newcommand{\hmpc}{\ifmmode{h^{-1}\,\hbox{Mpc}}\else{$h^{-1}$\thinspace Mpc}\fi}

\newcommand{\kms}{\ifmmode{\,\hbox{km\,s}^{-1}}\else {\rm\,km\,s$^{-1}$}\fi}
\newcommand{\msun}{{\rm\,M_\odot}}

\begin{document}
\title{Density Variations in the NW Star Stream of M31\altaffilmark{1}}
\shorttitle{M31 NW Star Stream}
\shortauthors{Carlberg and PAndAS}
\author{R. G. Carlberg.\altaffilmark{2} }
\author{Harvey B. Richer\altaffilmark{3}}
\author{Alan W. McConnachie \altaffilmark{4}}
\author{Mike Irwin\altaffilmark{5}}
\author{Rodrigo A. Ibata\altaffilmark{6}}
\author{Aaron L. Dotter\altaffilmark{7}}
\author{Scott Chapman\altaffilmark{5}}
\author{Mark Fardal\altaffilmark{8}}
\author{A. M. N. Ferguson\altaffilmark{9}}
\author{G. F. Lewis\altaffilmark{10}}   
\author{Julio F. Navarro\altaffilmark{11}}
\author{Thomas H. Puzia\altaffilmark{4}}
\author{David Valls-Gabaud\altaffilmark{12}}

\altaffiltext{1}{Based on observations obtained with MegaPrime / MegaCam, a joint project of CFHT and CEA/DAPNIA, at the Canada-France-Hawaii Telescope (CFHT) which is operated by the National Research Council (NRC) of Canada, the Institute National des Sciences de l'Univers of the Centre National de la Recherche Scientifique of France, and the University of Hawaii. }
\altaffiltext{2}{Department of Astronomy and Astrophysics, University  of Toronto, Toronto, ON M5S 3H4, Canada carlberg@astro.utoronto.ca }
\altaffiltext{3}{Department of Physics and Astronomy, University of British Columbia, Vancouver, BC V6T 1Z1, Canada richer@astro.ubc.ca}
\altaffiltext{4}{NRC Herzberg Institute for Astrophysics, 5071 West Saanich Road, Victoria, BC V9E 2E7 Canada alan.mcconnachie@nrc-cnrc.gc.ca,Thomas.Puzia@nrc.ca}
\altaffiltext{5}{Institute of Astronomy, Madingley Road, Cambridge, CB3 0HA, U.K. mike@ast.cam.ac.uk, schapman@ast.cam.ac.uk}
\altaffiltext{6}{Observatoire de Strasbourg, 11, rue de l'Universit\'e, F-67000, Strasbourg, France ibata@astro.u-strasbg.fr}
\altaffiltext{7}{Department of Physics and Astronomy, University of Victoria, Victoria, BC, V8P 1A1, Canada dotter@uvic.ca}
\altaffiltext{8}{Dept of Astronomy, University of Massachusetts, Amherst, MA 01003-9305, fardal@astro.umass.edu}
\altaffiltext{9}{University of Edinburgh, Blackford Hill, Edinburgh UK EH9 3HJ, ferguson@roe.ac.uk}
\altaffiltext{10}{Sydney Institute for Astronomy,  University of Sydney  NSW 2006, Australia  gfl@physics.usyd.edu.au}
\altaffiltext{11}{Department of Physics and Astronomy, University of Victoria, Victoria, BC V8P 5C2, Canada,  jfn@uvic.ca}
\altaffiltext{12}{GEPI, CNRS UMR 8111, Observatoire de Paris, 5 Place Jules Janssen, 92195 Meudon, France, david.valls-gabaud@obspm.fr}

\begin{abstract}
The Pan Andromeda Archeological Survey (PAndAS) CFHT Megaprime survey of the M31-M33 system has found a star stream which extends about 120 kpc NW from the center of M31.   The great length of the stream, and the likelihood that it does not significantly intersect the disk of M31, means that it is unusually well suited for a  measurement of stream gaps and clumps along its length as a test for the predicted thousands of dark matter sub-halos. The main result of this paper is that the density of the stream varies between zero and about three times the mean along its length on scales of 2 to 20 kpc. The probability that the variations are random fluctuations in the star density is less than $10^{-5}$.  As a control sample we search for density variations at precisely the same location in stars with metallicity higher than the stream, [Fe/H]=[0, -0.5] and find no variations above the expected shot noise. The lumpiness of the stream is not compatible with a low mass star stream in a smooth galactic potential, nor is it readily compatible with the disturbance caused by the visible M31 satellite galaxies.  The stream's density variations appear to be consistent with the effects of a large population of steep mass function dark matter sub-halos, such as found in LCDM simulations, acting on an approximately 10~Gyr old star stream. The effects of a single set of halo substructure realizations are shown for illustration, reserving a statistical comparison for another study.
\end{abstract}

\keywords{dark matter; Local Group; galaxies: dwarf}

\section{INTRODUCTION}
\nobreak
Galactic halos formed from LCDM initial conditions in n-body simulations have approximately 10\% of their mass in orbiting sub-halos \citep{VL1,Aquarius}. The gravitational stirring and heating of  a galactic disk and halo star clusters by dark sub-halos is tempered by the large numbers, high random velocities and broad distribution of sub-halos throughout the  overall dark halo.  However, very low velocity dispersion star streams in galactic halos are sensitive to the degree to which the dark matter in the halo is sub-structured into thousands of orbiting sub-halos. The sub-halos fold and chop the star-streams and gradually increase the velocity dispersion around the mean motion to about $\simeq$15\% of the halo circular velocity, typically 30\kms, over a Hubble time \citep{Ibata:02,SGV:08,StarStreams,YJH:10}.  Consequently,  cool star streams are sensitive indicators of the presence of the predicted dark matter substructure.  

About half a dozen of the currently known Milky Way streams qualify as cool (most confidently, Pal 5, GD-1, Orphan, Archeron and Styx) that is, having local velocity dispersions below about 10\kms, or, width less than about 0.1 radian as seen from the center of the host galaxy.   A dark matter sub-halo crossing such a cool stream will lead to visible disturbances. Unfortunately, the star count data often do not yet have sufficient local numbers to allow statistically significant measurements of density variations relative to the galactic foreground and background \citep{Odenkirchen:02, Ferguson:02,Majewski:04,Chapman:08, Grillmair:09,Odenkirchen:09}. 

The Pan Andromeda Archeological Survey (PAndAS) \citep{Pandas} in one fell swoop provides deep and uniform data to a distance of about 150 kpc from M31's center. The spectacular star stream north-west of M31, more than 100 kpc long, was first displayed in its entirety in \citet{Richardson:11}. The great length of the stream, as well as being fairly distant from the disturbing effects of the main body of M31, make the stream an exceptionally interesting case for analysis of density variations. And, the star counts have sufficient signal to noise to allow reliable local surface density measurements of the stream.

Multiple image gravitational lensing by a galaxy of a background quasars is another probe of sub-halos. 
\cite{MS:98} pointed out that the anomalous flux ratios relative to a smooth potential model that accurately predicts the locations of the images are the natural consequence of substructure in galaxy halos.  A statistical analysis of available strong lens systems \citep{DK:02,KD:04} assuming NFW density profile sub-halo profiles found that about 2\% of the halo mass (with a very large uncertainty)  is in substructure.  Lensing is very sensitive to the central density profile which \citet{Aquarius} found to be close to an Einasto profile, which is shallower than the NFW profile. 
 The theoretical situation has become somewhat unclear, since a projection of a sub-halo rich n-body galaxy simulation \citep{Xu:09,Xu:10} finds image distortions generally smaller than those seen in strong lens systems.  A better overall understanding and tighter limits requires better data on more strong lens systems \citep{Treu:10}.

\begin{figure}
\begin{center}
\includegraphics[bb= 10 0 750 500,width=0.75\textwidth]{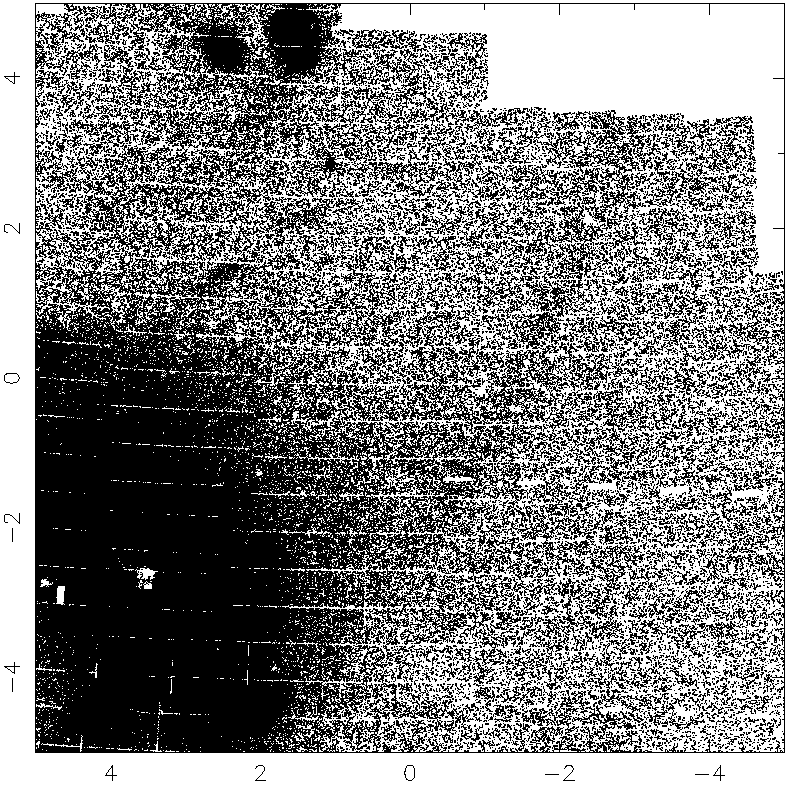}
\end{center}
\caption{The spatial distribution of the [Fe/H]= [-0.6, -2.4] red giant stars in the NW region of M31.  A full field version is presented in \citet{Richardson:11} The image is 10\degr\ across in the tangent projection co-ordinates, which are centered in the exact middle of this map. }
\label{fig_M31}
\end{figure}

In this paper we analyze M31's NW star stream for the density variations that sub-halos are expected to induce.  We first trace the mean centerline of the stream. The luminosity function and total luminosity of the main part of the stream are estimated as an indication of the progenitor system. The stream's width and deviations from the centerline are measured. The measurement of the density along the stream gives the data which we use to test whether the stream is smooth or lumpy. 
An illustrative simulation is used to show  the expected scale and degree of stream structure as the amount of substructure increases from the visible dwarfs to the thousands of halos that LCDM predicts.

\section{A MAP OF THE M31 STELLAR HALO}
\nobreak

PAndAS has acquired $g$ and $i$ band images covering nearly 400 square degrees of M31-M33 with the CFHT Megaprime system. The images have a minimum S/N=10 at $i_{AB}\simeq 24.5$ mag.   Details of data reduction are in \citet{McC:10} and \citet{Richardson:11}. We adopt the  \citet{McConnachie:05} distance of  785 kpc (DM of 24.47 mag) for M31, although we note a statistically consistent, but somewhat smaller, Cepheid distance  having a DM 24.32$\pm0.12$ mag or 731 kpc \citep{VJR:07}. The adopted distance indicates that the survey reaches to M$_i\sim  0$ mag at high signal-to-noise.  We  use data up to and including January, 2010.

\begin{figure}
\includegraphics[bb= 0 0 450 500,width=0.45\textwidth]{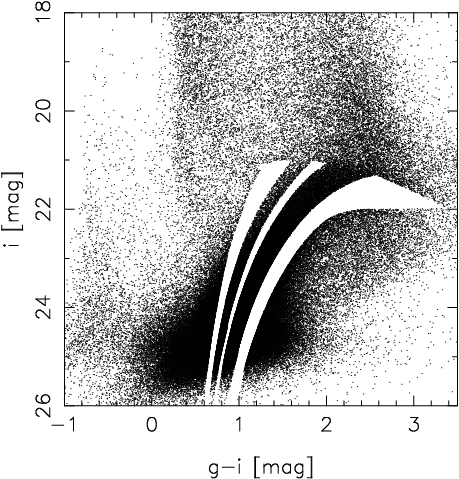}
\caption{The  color-magnitude diagram of stars in the M31 field. The stars to which the \citet{Dotter:08} isochrones assign metallicities, under the assumption that they are red giants at the distance of M31, are masked out at [Fe/H]= [0, -0.5], [-0.9, -1.1] and [-1.5, -3], from right to left, respectively.} 
\label{fig_cm}
\end{figure}

Image extraction yields approximately 31 million stellar objects in the images. Matching the colors and magnitudes of the stars to isochrones of varying metallicity at a fixed age of 12~Gyr \citep{Dotter:08} identifies M31 Red Giant Branch (RGB) stars using the procedures of \citet{McC:10}.  Since we are seeing M31 through our own galaxy, it turns out that more than 80\% of the stars do not match RGB isochrones at the adopted distance. These stars are not included in the map, although all stars above the completeness limit are useful for the measurement of geometric biases of the images.  The stars are individually extinction corrected with galactic dust maps \citep{McC:10}. We  are left with about 4.3 million nominal red giant branch stars to 24.5 mag which have been  assigned metallicities in the range of [Fe/H] =[-3, 0]. Of these stars, approximately 25\% are in the NW quadrant which is our area of interest. The galactic stars that happen to fall into the range of colors and magnitudes that are appropriate for M31 red giants create a slowly varying foreground which will be removed in the analysis of the stream density. Note that a shorter distance would make the inferred luminosities smaller and cause the isochrone match to systematically yield larger metallicities. To a first approximation this simply offsets the metallicity range we analyze and does not otherwise affect the analysis.

Star streams are identified as spatially coherent over-densities,  over a limited range of metallicities. To create the map we project the RGB stars onto a tangent plane, for which we chose a center at RA=0h 24 m and Dec = +44\degr\ which is near the center of the NW stream. This center is offset from M31's center, to make the co-ordinates reasonably rectangular in the region of the stream analysis.    Without the RGB star selection no streams are readily visible in the halo.  
Figure~\ref{fig_M31} plots the sky distribution (in the tangent plane co-ordinates) of the [Fe/H]=[-0.6, -2.4] stars which we find below comprise the stream. The same feature is clear in any map made with the low metallicity star sample.

The low metallicity map, Figure~\ref{fig_M31}, has a number of geometric sampling variations that must be controlled to make a reliable measurement of local  stream density. The simplest problem is that the CCD array has gaps between the individual chips of $\sim$80\arcsec\ in the horizontal direction and $\sim$13\arcsec\ in the vertical direction. The images were dithered to cover the smaller gaps,  but the resulting depth is shallower than the surrounding area and no simple count correction procedure that did not increase the noise was found. These gaps are identified in the camera CCD co-ordinate system and masked out of the map. A more complicated geometric bias results from the catalog being built up from stacked local images in which the stellar photometry was done to create a local catalog. These catalogs were then matched in the areas of overlap to eliminate duplicates and produce a global catalog. The images are vignetted towards the edge of the array so the image depth drops at the field edge,  however the vignetting is partially compensated in the catalog matching at the image edges. That is, if the surface density is corrected with the inverse of the mean star density over the array then the map has an excess at the edges. It is straightforward to devise a  correction which flattens the average surface density. The correction works well over a range of about 20\% in surface density, but undesirably amplifies the noise if applied at lower completeness levels. We apply a cut to the image where the local surface density in CCD image co-ordinates drops below 20\% of the average. The resulting masked out area is about 15\% of the image, a fraction which is not very sensitive to the precise value of the cut. The masked out regions are not included in the analysis and are readily visible in Figure~\ref{fig_M31}.  

\begin{figure}
\includegraphics[bb= 10 10 900 600,width=0.75\textwidth]{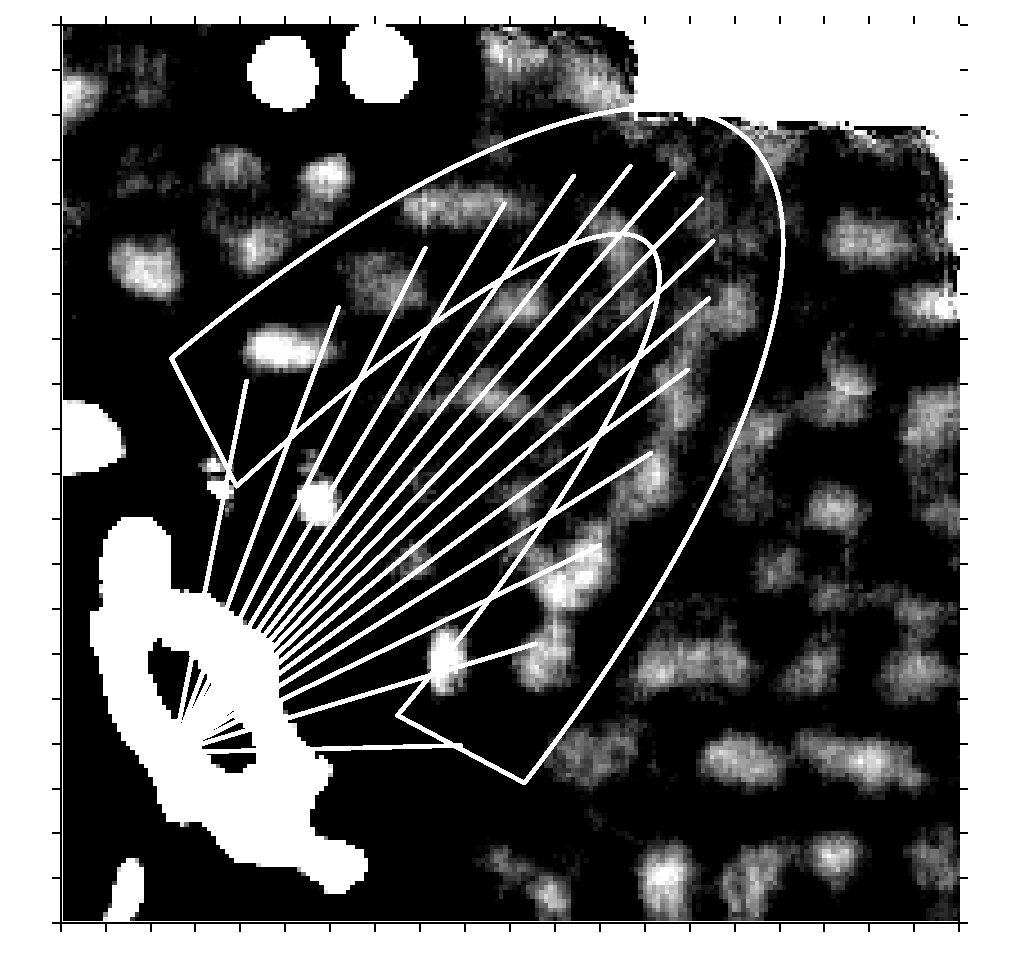}
\caption{A  Mexican top-hat filtered image (inner radius 0.4\degr\ and outer 0.8\degr) that emphasizes the structure, but is not used for analysis. The image covers the same area as Fig.~\ref{fig_M31}, with tick marks at 0.5\degr. The full width of the region used in the analysis, of which the outer half is used to define the local background, is shown.  The gray scale extends from 0 to 3 RGB stars per kpc$^2$, greatly emphasizing the contrast in the lumps.  The elliptical angles, defined from the $r(\phi)=a\cos{\phi}+b\sin{\phi}$ fit once rotated and centered, begin at $\phi$=-75\degr\ (nearly horizontal line from the center) and proceed counter-clockwise in increments of 10\degr\ to $\phi$=+70\degr\ (nearly vertical line). Note that these angles are not equal to the geometric angle. For our density variation measurements we use the well defined lower branch of the stream $\phi=[-70,-10]$. }
\label{fig_cutout}
\end{figure}

Field to field photometric depth variations will introduce artificial variations in the density of stars in the map.  The depth of an image for a fixed exposure time depends on the sky transparency, brightness and image quality at the time of observation.  Below the completeness limit there is substantial field to field variation that leads to a random checkerboard appearance, but the pattern essentially disappears above the completeness limit. Our measurement of the stream density subtracts the local mean background to reduce the problems of local depth variations.
  
A second, more difficult, form of depth variations is field to field variations in the zero points of the two filter bands.  We can set a limit on these by using the double measurements of star brightness in the overlap regions.  For bright stars where the photon noise is small the standard deviation is 0.037 mag in $g$ band and 0.052 mag in $i$ band, with no discernible pattern over the field. The star catalog in our selected range of color and magnitude has nearly constant numbers with increasing depth near the completeness limit, so a 5\% change in zero point leads to about a 5\% change in numbers.  The 5\% should be considered an upper limit since the edges of the images are vignetted and have significant local calibration variations. Across each 1\degr\ Megaprime field there are a myriad of camera co-ordinate dependent photometric offsets \citep{Regnault:09} which the current calibration does not fully take into account. These offsets lead to a change in zero point within each one degree pointing that is an additional $\sim$2-3\% variation.  The masking procedure removes the worst of these variations, and the non-linear correction to flat star counts substantially reduces the variations within a field. What remains is below the shot noise of the star counts.

\subsection{The North-West Stream}

The North-West stream, readily visible in the star map of Figure~\ref{fig_M31}, extends more than 100 kpc from M31 in projection.   An ellipse is used to trace the stream, which corresponds to finding the lowest order Fourier mode, that is, $r(\phi)=a\cos(\phi)+b\sin(\phi)$, rotated and centered for a best fit. Note that the angle $\phi$ is the angle once the ellipse is transformed into a sphere and is not the geometric angle except along the major and minor axes. Higher order Fourier terms are not yet justified given the signal to noise of the stream within these data.  The ellipse parameters are varied, along with the stream width and the background region width to maximize the mean over-density within $\pm0.4\degr$ of the ellipse over the lower branch of the stream, taken as the angular range $\phi=[-70,-10]$. The stream region is shown in the Mexican top-hat filtered  Figure~\ref{fig_cutout}. This filtered image, with an inner positive region of radius 0.4\degr\ and outer negative region of radius of  0.8\degr, with an area integral of zero,  usefully illustrates the stream and noise properties near the filter length. The analysis is done on an unfiltered high resolution version of Figure~\ref{fig_M31}  with pixels of 18\arcsec, ten times smaller than in Figures~\ref{fig_M31} and \ref{fig_cutout}.

We find that the best fit ellipse has a location of [-1.25\degr,+1.90\degr] in the tangent plane co-ordinates of Figure~\ref{fig_M31} relative to our field center  at [0h 24m,+44\degr]. There is no particular reason for this fitting procedure to be constrained to have an ellipse centered on M31. The ellipse has axes of $8.55\degr\ \times 2.40\degr$ (or $117 \times 33$ \kpc) at an angle of 42.25\degr\ W of N.  The best fit occurs for an ellipse that is within the image data rather than for one that extends off the imaged area, even if the fitting procedure is started with a longer ellipse that extends out of the field.  

\begin{table}
\begin{center}
\caption{[Fe/H] dependence of mean surface density}
\label{tab_slices}
\begin{tabular}{rr}
\tableline\tableline 
{\rm [Fe/H] --} & $\overline{\mu}$ \\
{\rm [Fe/H] - 0.2} & stars kpc$^{-2}$  \\  
\tableline 
-0.60 &     0.05 \\
-0.80 &     0.16 \\
-1.00 &     0.41 \\
-1.20 &     0.53 \\
-1.40 &     0.27 \\
-1.60 &     0.28 \\
-1.80 &     0.06 \\
-2.00 &     0.17 \\
-2.20 &     0.08 \\
-2.40 &     0.01 \\
\tableline
\end{tabular}
\end{center}
\end{table}

The straightforward consequences of projection are that the stream distance from M31 must be 117 kpc or larger, and, the ellipse must be at least as circular as we found above. We have assumed that all the stars in the stream are at 785 kpc, the adopted distance to M31. However, the distances of stars around the stream could vary roughly 100 kpc closer or further away, depending on how the stream is projected.  Such distance variations will cause a distance modulus error of nearly $\pm0.3$ mag, which means that individual stars will be identified with isochrones of the incorrect metallicity.  This would be a significant problem if the stream were a narrow range in metallicity. However, our unweighted selection of stars over the very broad range of [Fe/H]=[-0.6, -2.4] corresponds to selecting a large rhombus in color-magnitude space that is 3.1 mag high in absolute magnitude.  The upper limit to the distance differences of the stream relative to the body of M31 would blur out our selection of stream stars about 10\% or so, but it is only a small dilution of the stream density.  

\subsection{Structure perpendicular to the Stream}

\begin{figure}
\includegraphics[bb= 10 10 800 500,width=0.75\textwidth]{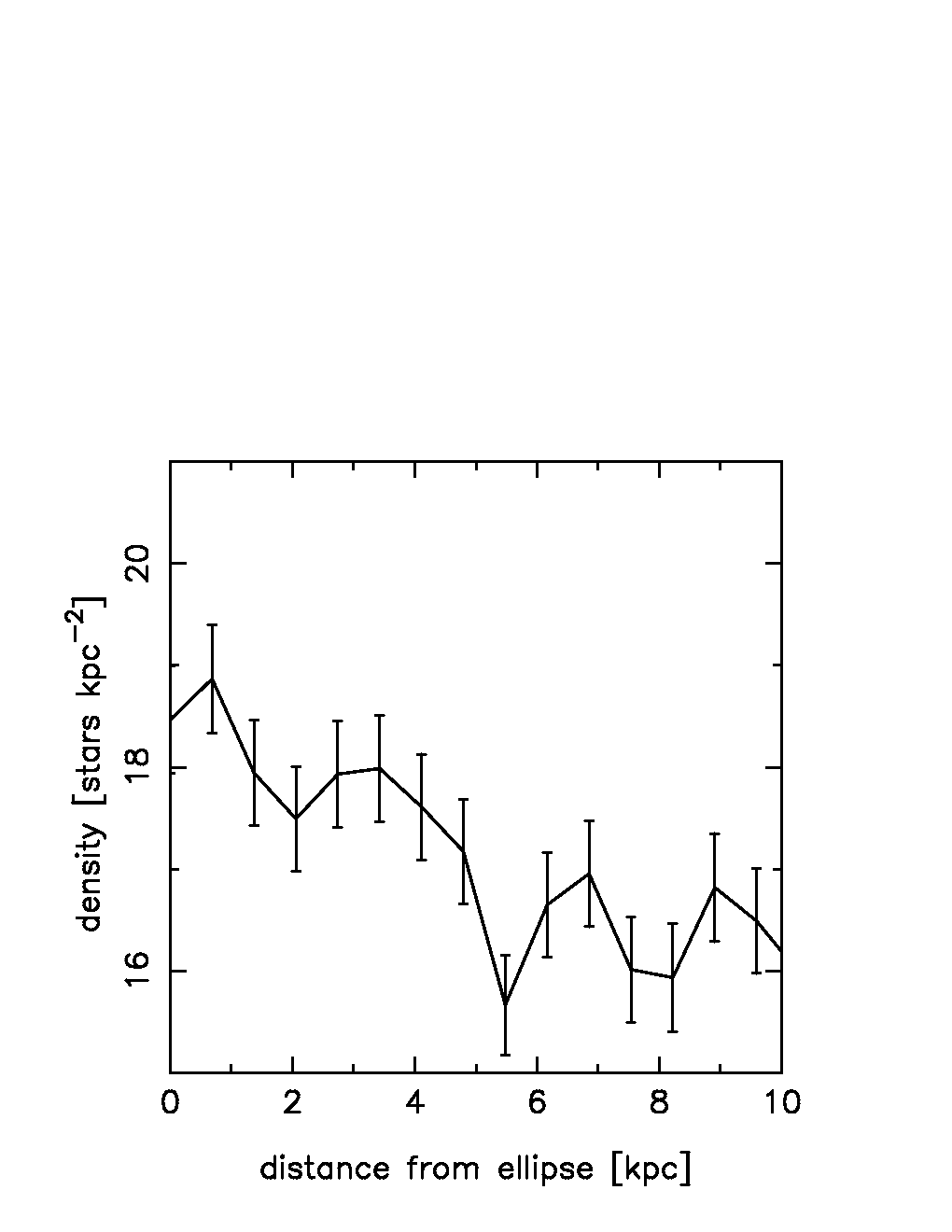}
\caption{The mean density transverse to the best elliptical fit to the stream over the $\phi=[-70,-10]$ range. Errors are estimated from $\sqrt{N}$ in the bin.  }
\label{fig_stream_width}
\end{figure}

We measure the density perpendicular to the ellipse ridge line and display the result in Figure~\ref{fig_stream_width}. The stream has a half width at half of the peak density of about 2.5~kpc or a full width of 5~kpc.   The stream width is dominated by the physical width of the stream, but also includes a significant component from the stream location deviating slightly from our assumed elliptical shape.  The stream subtends an angle ranging from about 0.1 radian to 0.05 radian, as seen from the center of M31, although the stream is physically wider than the cool, but relatively close, streams in our own galaxy.

\begin{figure}
\includegraphics[bb= 10 10 800 500,width=0.75\textwidth]{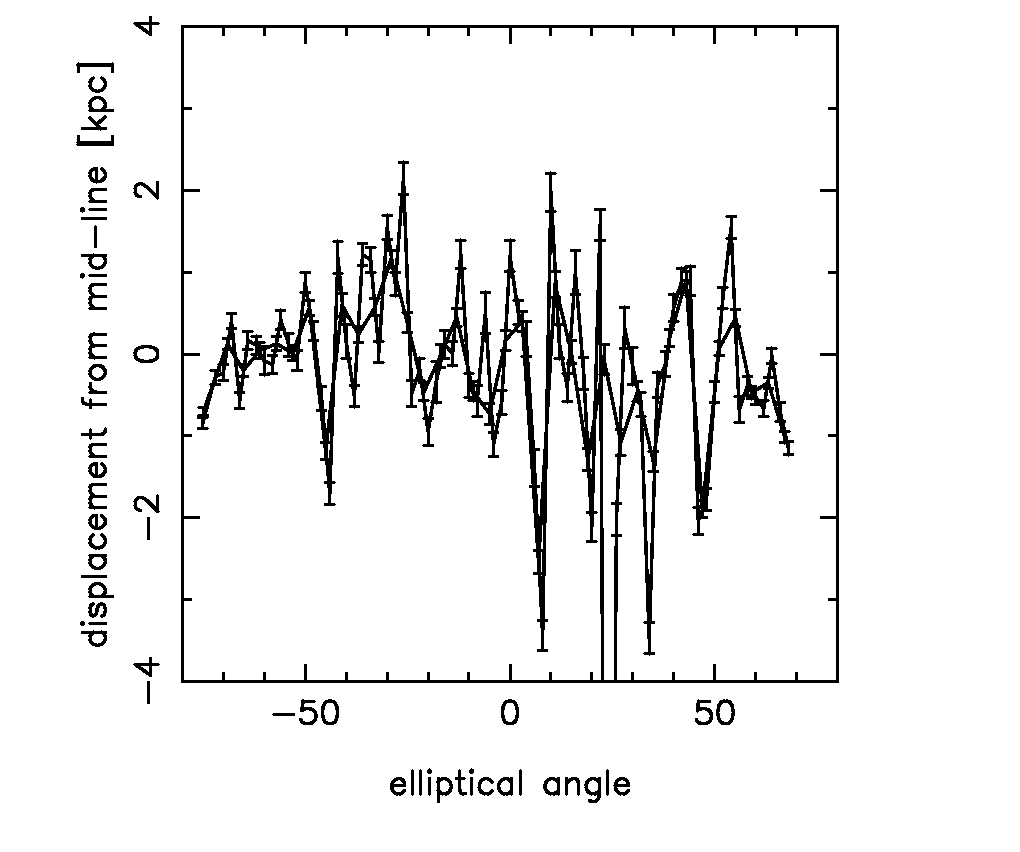}
\caption{ The location of the stream center relative to the best fit ellipse centerline, for 2\degr\ and 4\degr\ binning. The variations are larger than the expected statistical variations, particularly for the "tilted" segments at positive angles.}
\label{fig_stream_center}
\end{figure}

The location of the centerline of the stream, calculated as the density weighted mean distance from the best fit ellipse, is shown in Figure~\ref{fig_stream_center}. The error flags are calculated from the statistics of the star counts and confirms the impression of Figure~\ref{fig_cutout} that the stream center varies, sometimes discontinuously, from the best fit ellipse.  The variations are largest on the upper branch of the stream, $\phi>0$, where it has large gaps. On the more continuous lower branch there is  a statistically significant offset from the mean centerline at $\phi\simeq -30\degr$.

\subsection{Metallicity and Luminosity Distribution of the Stream}

The density is measured within a constant width region around the ellipse, $\pm 0.4\degr$ ($\sim5$~kpc).  We subtract the foreground/background measured at the same angular bin within the two adjacent 0.2\degr\ strips on either side of the stream.  The range of metallicity in the stream is shown in Table~\ref{tab_slices} in bins of  width 0.2 in [Fe/H]. Table~\ref{tab_slices} gives the top of the metallicity bin in column 1, the mean number of stars per kpc$^2$ in column 2. This table shows that the stream has its maximum density in the [Fe/H]$\sim -1.2$ bin.  For our measurement of the stream density we will use the entire [Fe/H]=[-0.6,-2.4] range, uniformly weighted. In principle somewhat better signal to noise would be obtained if we weighted the stars with the metallicity distribution function. However, the flat distribution gives the most straightforward stream density measurement. 

The stream is broadly distributed in brightness with the mean number of stars [21-22, 22-23, 23-24, 2$\times$(24-24.5)] $i$-band mag in the [Fe/H]=[-0.6, -2.4] range being  [0.62, 0.78, 0.55, 0.47] per magnitude per  kpc$^2$ over the -70 to -10 angle range. The luminosity function of the stream is essentially flat over the absolute magnitude range $M_i \le +0$ as is usually the case for low metallicity RGB luminosity functions.   Physical properties of the stream are derived below.

\section{DENSITY STRUCTURE OF THE STREAM}

Whether the NW stream is smooth or lumpy is the observational question of primary interest in this paper. This measurement needs to be handled carefully since the stream over-density is only about 15\% of the total nominal RGB star density.  The stream density will vary simply due to the $\sqrt{N}$ fluctuations in the few thousand stars in each angular bin.  The density measurement is done within the region shown in Figure~\ref{fig_cutout}. The stream is defined as being a fixed width of the centerline and the local background is measured within the cutout region beyond the stream. The analysis region of both the stream and local background is shown in Figure~\ref{fig_cutout}.  The newly identified dwarf galaxy AndXXVII \citep{Richardson:11}, just above the upper branch of the stream in Figure~\ref{fig_M31}, is masked out to exclude it from the local background correction.

\begin{figure}
\includegraphics[bb= 10 10 800 500,width=0.75\textwidth]{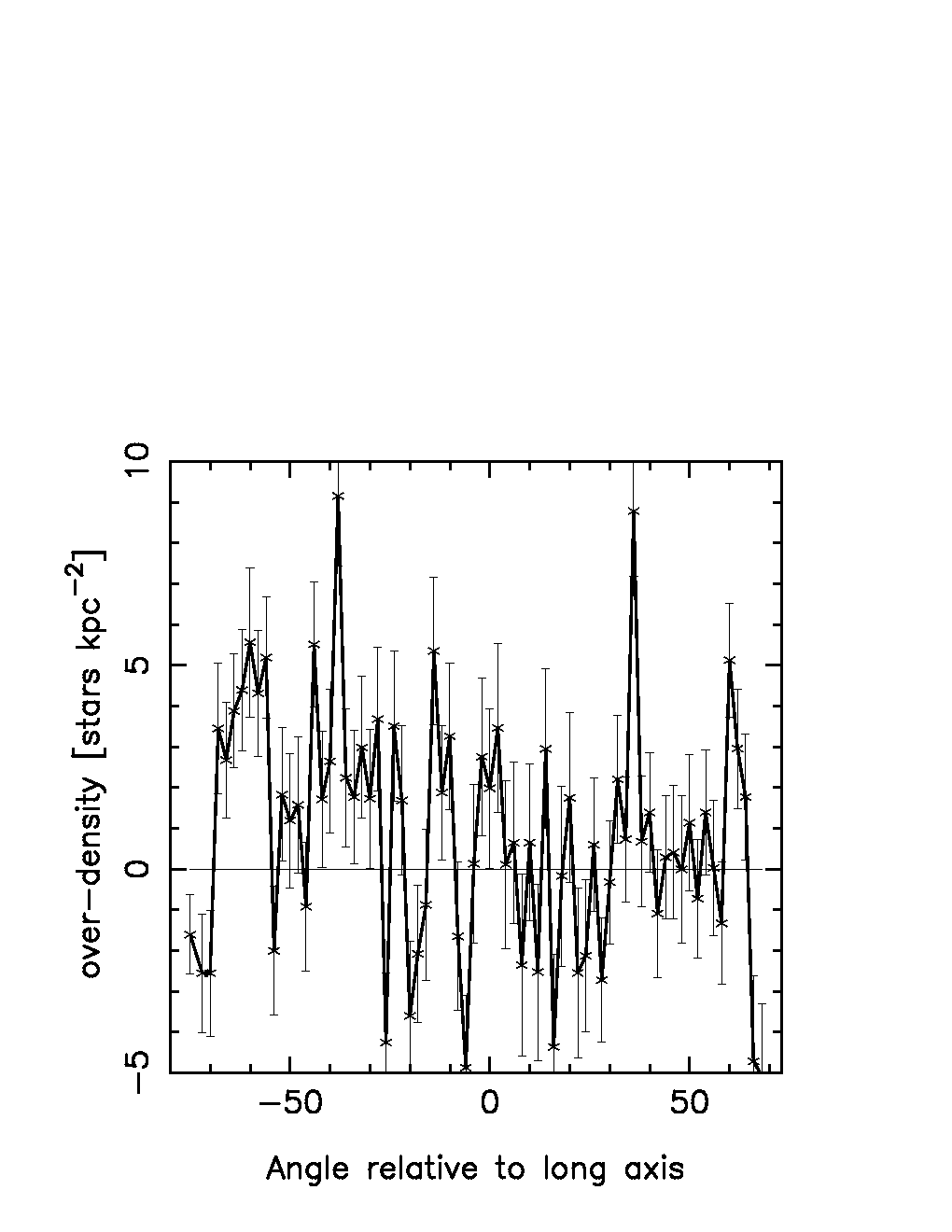}
\caption{The mean density around the stream in bins of 2\degr\ of elliptical angle. The errors are estimated from the star counts in the bins and the surrounding background region. The $\chi^2$ statistics is calculated only for the $\phi=[-70,-10]$ range. }
\label{fig_stream_density}
\end{figure}

The density along the stream, within a half-width of 0.4\degr\ on the sky subtracting the local background in the adjacent 0.2\degr\ regions, is shown in Figure~\ref{fig_stream_density} at intervals of 2\degr\  in the ellipse angle, $\phi$. The errors are computed from the $\sqrt{N}$ of the star-counts in the stream and in the background region, added in quadrature.

\begin{figure}
\includegraphics[bb= 10 10 900 600,width=0.75\textwidth]{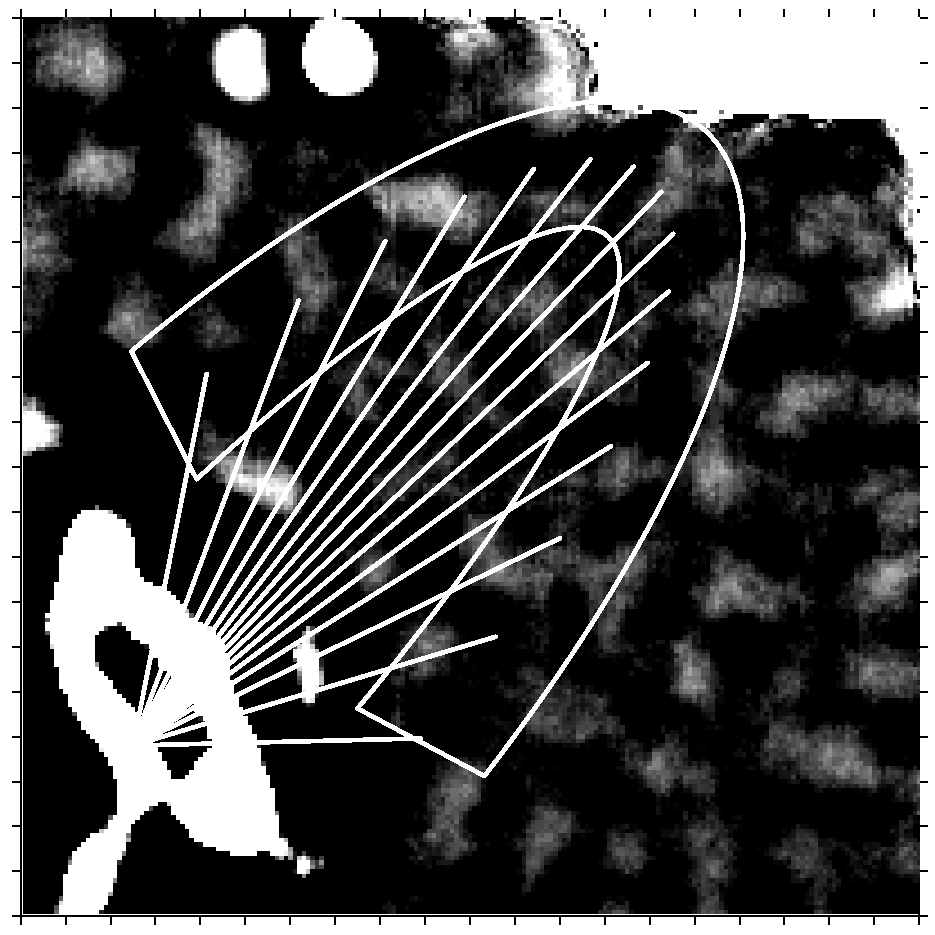}
\caption{Same as Fig.~\ref{fig_cutout} showing that the stream is not evident in the [Fe/H] = [-0.5, 0]. The contrast scale is the same as in Fig.~\ref{fig_cutout}.}
\label{fig_Zhi}
\end{figure}

\subsection{Statistical Significance of Density Variations}

The $\chi^2$ test estimates the probability that a set of data is consistent with a model, which in our case is a constant density.   We  need to be aware that sampling and calibration variations could artificially create density variations. Masking takes care of the CCD gaps and the low exposure regions of the images. The field flattening largely corrects the counts over the rest of the image. Any residual photometric fluctuations will be measured in the control fields to estimate their size.

The $\chi^2$ statistic applied to the $\phi=[-70,-10]$ range indicates that the chance that the stream has a uniform density is less than about $10^{-6}$.  A binning angle of 2\degr\ corresponds to roughly 2-4 kpc, depending on location around the ellipse. We have measured the stream density variations in bins ranging from 1\degr\ to 8\degr\ finding that the variations are present at all these scales. Since there are easily  visible clumps and gaps for $\phi \ge -10$ including the upper branch of the stream in the $\chi^2$ analysis would significantly increase $\chi^2$, which of course could well be a real description of the larger stream. We prefer to present the most conservative statistic on the visibly smoothest part of the stream. We emphasize that even after restricting our statistical analysis to the lower branch, the segment is the longest stream segment available for study so far. Moreover its orbit appears to keep it well away from the potential disturbances of the structure in the disk of M31. 

\begin{figure}
\includegraphics[bb= 10 10 800 500,width=0.75\textwidth]{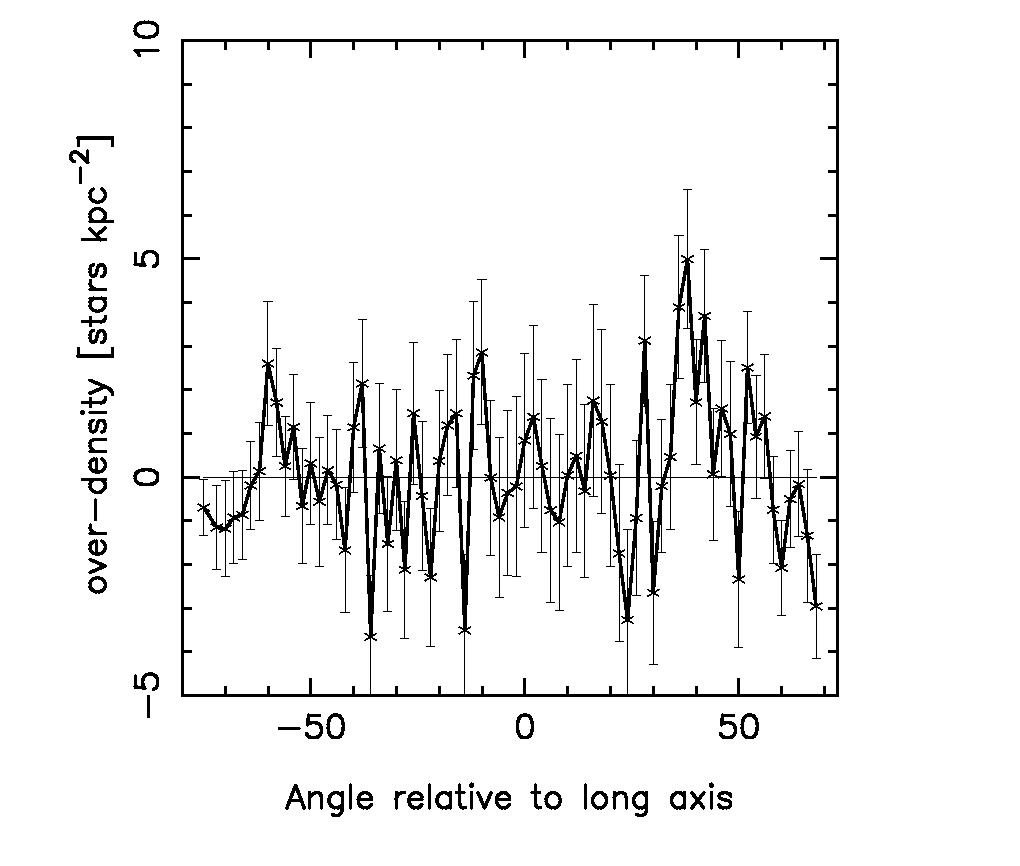}
\caption{ The angular dependence of the stream density at high metallicity, showing that the density variations due to background are not significant. The error flags at each point are about 30\% smaller than for the stream because there are fewer stars.}
\label{fig_density_Zhi}
\end{figure}

\begin{table}
\begin{center}
\caption{Probability of a Smooth Stream}
\label{tab_chi2}
\begin{tabular}{rrrc}
\tableline\tableline 
  & & \\
$\Delta\phi$ & $\chi^2/\nu$ & $\nu$ & $1-P(\chi^2,\nu)$ \\
 \\
\tableline
\multicolumn{4}{c}{NW Stream} \\
 1  &2.18 &  59 & $4\times10^{-7}$\\
 2  &3.01 &  28 & $2\times 10^{-7}$\\
 4  &3.58 &  14 & $6\times 10^{-6}$\\
 \\
\multicolumn{4}{c}{Null test with [Fe/H]$\ge$-0.5} \\
 1  &1.17 &  59 & 0.17\\
 2  &1.16 &  28 & 0.25\\
 4  &0.65 &  14 & 0.85 \\
 \\
 \multicolumn{4}{c}{$2\degr \times 4\degr$ offset stream } \\
 1 & 1.96 &  29 & 0.002 \\
 2 & 1.57 &  13 & 0.090 \\
 4 & 2.38 &   6 & 0.027\\
\tableline
\end{tabular}
\end{center}
\end{table}

We have varied virtually all stream and image parameters to check the result that the density variations are truly those of the star stream.  Reasonable changes in the path of the ellipse, its width and the local background make little difference, other than slightly reducing the mean density of the stream.  The photometric zero point variations that remain in this version of the data are readily estimated.  The mean background is approximately 12 stars kpc$^{-2}$ and the photometric variations are expected to produce 5\% or smaller fluctuations coherent over scales of 1 degree in the image, or a standard deviation of about 0.6 stars kpc$^{-2}$.  The measured density fluctuations have a standard deviation of $\simeq$2.0 stars kpc$^{-2}$, at least three times the size of the variations that the photometric zero points are expected to cause.

A straightforward control sample is to measure the fluctuations at the same location in higher metallicity stars, [Fe/H]$\ge$-0.5, in which the stream is not detected. Figure~\ref{fig_Zhi} shows the Mexican top-hat filtered image in the high metallicity stars, demonstrating that the background fluctuations have a significantly different pattern but have a similar amplitude.  The density around the stream is displayed in Figure~\ref{fig_density_Zhi}. The $\chi^2$ test for the significance of the density variations is reported in Table~\ref{tab_chi2}.   No significant variations are found.  

A second control sample is to simply offset the measurement region away from the stream. We add 2\degr\ to the minor axis and 4\degr\ to the major axis and repeat the measurements, adjusting the $\phi$ range to approximately the same distance. The statistics are reported in Table~\ref{tab_chi2} finding that the probability that the variations are due to chance to be about 5\%. Moving the region around find finds that the $\chi^2$ values fluctuate a fair bit depending on position although nothing as significant as the NW stream itself ever emerges. These fluctuations are likely the result of shorter and lower significance star streams that are real entities. 

\subsection{Upper Branch Stream Substructure}

Whether the upper branch of the stream in Fig.~\ref{fig_M31} or \ref{fig_cutout} is the extension of the lower branch of stream is unclear in these data. The upper branch certainly fits within the same ellipse and has a shared metallicity distribution at the precision we can measure it. However, the upper branch is less a stream than three dominant segments separated by gaps of comparable size. It is of great interest to note that the segments are tilted with respect to the general path of the stream. Simulations show that tilted segments are some of the dominant structures frequently seen in chopped streams, as first noted in \citet{YJH:10}, and as we show below.  

\subsection{Dynamics and Alternate Clumping Mechanisms}

The total luminosity of the lower branch of the stream is found by summing the over-densities in one magnitude bins within a range of  [-70,-10] ecliptic angle, to find a stream segment luminosity of  $4.3\times 10^5 \lsun$ to $M_i\simeq 0$ mag. The luminosity is the sum over the RGB stars alone. To estimate the total luminosity we use the cumulative stellar luminosity function of a stellar population of similar age and metallicity, for which we chose the globular cluster M12.  M12 is an intermediate metallicity cluster with a well studied stellar luminosity function \citep{Hargis:04}.  Using their distance modulus of 14.0 mag and their I band luminosity function we find about 58\% of the light above $M_i=0$ mag.
 Therefore the corrected total luminosity is $7.4\times 10^5 \lsun$. 

The mean surface brightness is $1060\lsun\, \kpc^{-2}$.  If we assume the stream is a uniform density cylinder of 5 kpc radius and a stellar mass-to-light ratio of, say, 3 in the $i$-band, then the volume stellar mass density is $500 \msun\, \kpc^{-3}$.  This can be compared to the mass density in the dark matter halo at say 90 kpc, which for our adopted halo parameters is $1.2\times 10^5\msun\, \kpc^{-3}$, that is, about a factor of about 240 higher than the stellar stream mass density.  At 30 kpc the ratio is 3600 times more background halo dark matter than stream stellar mass. If dark matter from the progenitor is mixed into the stream then the mass density of the stream would be proportionally increased, lowering the local mass ratios but the stream is unlikely to be significantly self-gravitating anywhere.

\citet{QC:10} have raised the possibility that Jeans instabilities can be the source of of density variations in the stream. The Jeans length in the stream, $\lambda_J=\sqrt{\pi \sigma^2/G\rho}$ evaluated with the stellar mass density and assuming an internal velocity dispersion of 10 \kms\ is about 400 kpc. On the other hand if the velocity dispersion were as low as 1 \kms\ then the formal Jeans length would be 40 kpc, comparable to the longest scale of density variations in the stream. Very low velocity dispersions are expected for slowly tidally stripped globular clusters. The presence of several globular clusters along the course of the stream \citep{Mackey:10} and the possibility that AndXXVII \citep{Richardson:11} is the source of the stream suggests that the internal velocity dispersion is at the level that a dwarf galaxy would create, $\sim  3 \kms$ or more. The resulting Jeans length is then roughly 120 kpc, the size of the stream. It is important to note that the stability analysis is far from complete. First, the disrupted object may have its own dark matter which has identical kinematics to the visible stars and would increase the mass density. Second, the stability analysis needs to account for the very strong tidal field of the M31 dark halo. Strong tides generally act to suppress much instability, particularly in largely radially oriented streams, which is likely the case for the NW stream.

Another mechanism that generates clumps in tidal streams is simply the pileup of stars at low velocity points of their epicyclic orbits \citep{KMH:08}. The lumps appear at $12\pi$ times the tidal radius. Since no progenitor object for the stream has been confidently identified, the tidal radius is very uncertain. However if we take a low luminosity dwarf galaxy of stellar plus a relatively  low dark mass for a total progenitor dwarf mass of $M_d=10^8 \msun$, then at a galactic radius of $r_g=100$ kpc, the  tidal radius, $r_g(M_d/(3M_g))^{1/3}$,  is approximately 3.2 kpc. The epicycle pileup separation will then be at a separation of 120 kpc. That scale compares to the size of the stream. The stellar mass in the stream alone, $2\times 10^6 \msun$ for a fairly minimal assumed mass-to-light ratio of 3,  gives about 40 kpc epicyclic clumping scale, which is at the upper end of range of what we see, but still leaves the smaller scale variations to be explained. We conclude that epicyclic pile-up is not a significant clumping factor for the NW stream.

If the progenitor system had large internal density variations in a highly order velocity field they could be fed out into the stream to create lumps.  However, this idea runs into trouble with our astrophysical knowledge of low mass galaxies. Dwarf spheroidal galaxies are comprised largely of stars on randomly oriented orbits and have essentially no substructure, which when tidally disrupted would lead to a smooth stream.  Dwarf irregular galaxies do have substantial substructure in their gas and young stellar populations. The visible luminosity (corrected for light below our observational limit) in the lower branch of the stream is only $7\times 10^5 \lsun$. Dwarfs with such low luminosities usually have comparable rotational and random velocities which would quickly blur out stellar structures deposited into a stream. Given that \citet{Mackey:10} associate 3 or 4 globular clusters with the NW stream, which would lead to a unusually high ratio of globular cluster luminosity to progenitor luminosity, it it possible that the NW stream is only part of a larger stream that has yet to be clearly identified. The structures within star forming dwarf galaxies are often low mass stellar associations and clusters that fairly quickly dissolve as a result of stellar mass loss and their own internal dynamics. Spiral patterns are at best very chaotic in low mass dwarf irregulars.  Our assessment is that although carefully sub-structured young progenitor disk system could produce some of the stream lumpiness, appropriate astrophysical systems do not exist. 

\section{COOL STREAM SIMULATIONS}

Star streams normally originate from the tidal dissolution of either dwarf galaxies or globular clusters. In both cases tidal fields pull off outer stars or sometimes dissolve the entire system and distribute the stars in nearby orbits around the galaxy. The effect of adding randomly orbiting dark matter sub-halos on highly idealized star streams has been studied previously \citep{Ibata:02,SGV:08,StarStreams,YJH:10}, finding that the streams were folded, chopped and heated to about 30\kms\ over a Hubble time.  For the examination of the expected degree of lumpiness of star streams we need a more realistic model stream. The association of globular clusters with the NW stream \citep{Huxor:08,Mackey:10} indicates that a dwarf galaxy, likely  AndXXVII, is the source of the stream. 

The study of the effects of  dark matter sub-halos on the stream is inherently statistical since we are unsure to what degree the visible stars and gas give a complete census of the sub-halos, as highlighted in the celebrated missing satellite problem \citep{Klypin:99,Moore:99}. A detailed match to the orbit of the NW stream is not required for this very basic assessment of the action of sub-halos on a stream. The details of the stream creation process have relatively little to do with the subsequent sub-halo interactions with the stream. Our test particle stream originates from particles orbiting in a constant mass Plummer sphere  to which particles in the core region are gradually given velocities that boost them out into the region where tides can carry them away.  Future studies will report a full suite of self-gravitating simulations.

The simulation uses a galactic potential with a spherical NFW halo potential scaled to the results of the Aquarius simulation \citep{Aquarius}. The rotation curve of M31 is somewhat higher than the Milky-Way, but beyond 30 kpc mass modeling finds that the two galaxies are very similar \citep{Carignan:06}. For simplicity and comparability we therefore adopt the same disk-bulge potential for the disk and bulge as \citet{Johnston:98} which has a Miyamoto-Nagai disk potential \citep{MN:75} and a nuclear bulge modeled with a Plummer sphere.  The sub-halos are modeled as a collection of Plummer spheres following the distribution in masses, orbits and internal structure of the \citet{Aquarius} simulations. The star stream particles are evolved in the combined potentials of the background galaxy and orbiting sub-halos. We use time steps of 0.14~Myr, or about $10^5$ in a Hubble time, to ensure that the quickly moving particles are integrated properly over the small scale potentials of the sub-halos.

\begin{figure}
\begin{center}
\includegraphics[bb= 0 10 300 950,width=0.4\textwidth]{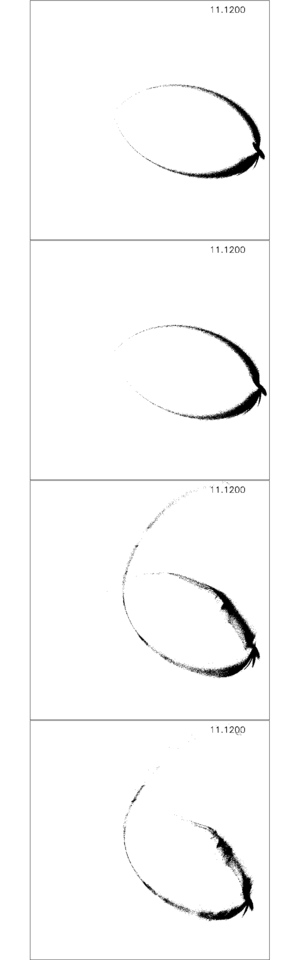}
\end{center}
\caption{ An example of stream disturbances under the action of 0, 5, 100 and 1000 dark matter sub-halos at an age of 11.12 Gyr. Note the increasing small scale variation as the number of sub-halos increases.}
\label{fig_Nsub_xy}
\end{figure}

\begin{figure}
\begin{center}
\includegraphics[bb= 0 10 300 950,width=0.4\textwidth]{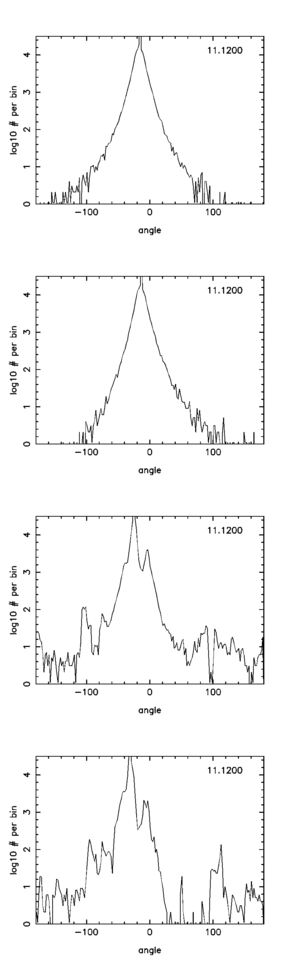}
\end{center}
\caption{ The density profiles in 2\degr\ angular bins around the streams of Figure~\ref{fig_Nsub_xy}.  Away from the progenitor at $\phi=0$ (which has remained bound in this case) the density variations approximately mimic the character of the M31 stream. Note that the stream is double valued around 90\degr.}
\label{fig_Nsub_den}
\end{figure}

\subsection{Density Variations in the Simulations}
The simulations start with the Plummer sphere $10^6 \msun$ progenitor located at $x=$ 100 kpc from the center in the plane of the disk, with $y=z=0$. The initial velocity is chosen to be entirely tangential perpendicular to the plane of the disk with an angular momentum of about 50\% of the circular value at that radius.   In the absence of sub-halos there are no density variations in the simulated star streams. 
M31 contains nearly 30 known dwarf galaxies. To what degree could these dwarfs, along with their dark-halos, cause the density variations we are measuring? For instance, what is the effect of the heaviest 5 sub-halos alone? Figure~\ref{fig_Nsub_xy} shows the $x-y$ plane projection of the stream in the presence of 0, 5, 100, and 1000 sub-halos as drawn from the $M^{-1.9}$ mass distribution. Including the 5 most massive sub-halos produces only large scale distortions of the stream relative to the smooth halo.  With 100 sub-halos we see significant small scale variations and going to 1000 sub-halos adds yet more small scale structure.  Figure~\ref{fig_Nsub_den} shows the density variations in angular bins assuming that the stream has a constant width, which is close to being equivalent to the M31 measurements shown in Fig.~\ref{fig_stream_density}.  In this particular simulation the progenitor remains bound, so 
 there is a density peak near $\phi=0$. Away from the progenitor there is a clear trend of increasing disturbance in the density variations as the number of sub-halos increases. Neither a smooth halo nor one with only the five most massive sub-halos, which roughly mimics the most massive dwarf galaxies, produce anywhere near the amount of substructure we see. One hundred sub-halos do a reasonable job in the outer parts  of the tail but do not do much damage to the inner part of the tail. Once we get to one thousand sub-halos we star seeing gaps and clumps essentially everywhere. 
 
Deviations from a mean centerline are also a signal of sub-halo interactions. Again the hundred halo simulation  produces a few  angled clumps but they are much more clearly present with the thousand halo simulation.  We will report on the statistics of a much larger set of simulations elsewhere. 

In this illustrative study we find that of order of one thousand sub-halos are required to produce all the small scale structure on the scales seen in the M31 NW stream from an initially smooth stream. 
\citet{YJH:10} report a more complete dynamical analysis focused on streams at smaller radii, but the dynamical outcomes they describe are quite generally applicable. They also firmly conclude that a large population of dark matter sub-halos will induce gaps and clumps in streams. A larger statistical study will be reported elsewhere.

\section{DENSITY STRUCTURE OF OTHER STREAMS}

A spectacular star stream in our own galaxy is the Pal~5 stream \citep{Odenkirchen:02} which is kinematically cold and narrow \citep{Odenkirchen:09}. At a distance of 23.2 kpc the visible part is only 4 kpc long. It is very narrow, about 0.4 kpc, roughly 1/5 of the width of the M31 NW stream segments. The source of the stream is the outer halo globular cluster Pal~5 with the stream having the highest surface density at the location of its progenitor. The density analysis of \citet{Odenkirchen:03} finds that there are density variations of the Pal 5 stream above the shot noise in the northern part of the stream, but not the southern part. The significance of the variation is not given, but there are 9 of 25 points more than $1\sigma$ from the mean trend over about 1 kpc of length. This certainly constitutes the detection of a significant lump in the currently available 4 kpc length of the Pal 5 stream, but on a smaller scale.

\citet{JSH:02} were inspired by the narrow velocity and geometric distribution of the 75 carbon stars of the Sagittarius stream \citep{Ibata:01} to consider the lumpiness of the Milky Way halo.  They develop a statistic based on the fifth through tenth Fourier terms to compare smooth halo models to models with up to 256 sub-halos, being careful to consider models that allow for the main dwarf galaxies. When applied to the available Sagittarius data they find that ``the degree of scattering is entirely consistent with debris perturbed by the LMC alone.''  If applied to the 100 kpc NW M31 stream, the tenth Fourier term would be a $\phi$=36\degr\ variation, which is roughly comparable to a 16\degr\ binning where there still is real structure relative to a constant, but only on yet larger scales.

\citet{Majewski:04} considered kinematic data for the Sagittarius stream in our galaxy and concluded that the halo could not be very lumpy, however we note that our simulations show that sub-halos tend to ``chop" streams leaving locally cold remnants with pieces at the same radius offset in velocity, which may be compatible with the \citet{Majewski:04} Figure~2. Kinematic tests will become very powerful in testing for substructure once large velocity samples are available.

\section{DISCUSSION AND CONCLUSIONS}

The 100 kpc NW stream in M31 is a coherent geometric structure whose orbit is well clear of the body of M31 making it a near ideal testing ground for the presence or absence of thousands of dark sub-halos predicted in LCDM n-body simulations.  Although the stream appears to be nearly a half ellipse over a common range of metallicities the upper branch of the stream is less well defined and has a number of clearly visible gaps. The lower branch is nearly complete and provides a much more conservative test for the presence of sub-halos.  The main result of this paper is that the stream has highly significant density variations on virtually all scales from 2 kpc, up to about 20 kpc.  The variation of density around the mean has a very low probability of being a chance statistical fluctuation, less than $10^{-5}$.  We have been careful to measure the lumpiness relative to an averaged local background and masked out gaps and regions where the photometric uncertainties add significantly to the variations.   As a control sample we take a higher metallicity set of stars at exactly the same location as the stream, which finds no significant density variations.  Relative to other known cool star streams, the M31 stream stands apart for its length, distance from the disturbing effects of the M31 disk, a variety of scales of substructure and its high statistical significance.

It is interesting to note that M31 appears to have only the one well defined long stellar stream at large radius. The NW stream has quite a low total luminosity so it should not particularly stand out relative to other stellar streams that were created at large radius over the buildup of the M31 dark halo.  Several other stream fragments are visible in the full field map \citet{Richardson:11} but none as long or coherent as the NW stream. If one assumed that other long streams likely formed over the lifetime of the halo then the absence of others at the present time indirectly suggests that they have been broken up by sub-halos \citep{StarStreams}.

The measured NW stream lumpiness essentially rules out the possibility that it is a low mass star stream orbiting in a smooth galactic potential of a disk plus bulge plus dark halo. Conversely, the NW stream density variations are compatible with the level of density changes that a large population of dark matter sub-halos induce.   The details of the density variation are sensitive to the statistical distribution of the sub-structure but are unlikely to be specifically modeled for a single stream. As more streams are observed statistical modeling of the degree of substructure present should become possible.  The present study is statistically consistent with a highly sub-structured dark halo, although we caution that the density data alone for a single stream is not a conclusive proof that halos are as sub-structured as LCDM simulations predict. Overall, the density variations of the NW stream are strong circumstantial evidence that the predicted thousands of dark matter sub-halos are present in M31's dark halo. 

\acknowledgements

This research is supported by NSERC, CIfAR and NRC in Canada and the French ANR programme POMMME.

\end{document}